\documentclass[11pt]{article}
\usepackage{amsmath}
\usepackage{graphicx}
\usepackage{amssymb}
\usepackage{amsfonts}
\usepackage{latexsym}

\def\beq{\begin{eqnarray}}
\def\eeq{\end{eqnarray}}

\def\al{\alpha}
\def\be{\beta}

\def\la{\lambda}

\def\om{\omega}

\def\La{\Lambda}

\begin{document}

\hfill {Preprint number: DF-UFJF-02-06-08}
\vskip 6mm

%
\begin{center}

{\Large\sc Polemic Notes On IR Perturbative Quantum Gravity}
\vskip 5mm

{\bf 
Ilya L. Shapiro}\footnote{E-mail address: shapiro@fisica.ufjf.br}
\vskip 4mm


Departamento de F\'{\i}sica -- ICE,
Universidade Federal de Juiz de Fora
\\
Juiz de Fora, CEP: 36036-330, MG,  Brazil
\vskip 2mm
\end{center}

\vskip 6mm

\begin{quotation}
\noindent
{\large\bf Abstract}\footnote{
Based on plenary talk at the SEVENTH ALEXANDER FRIEDMANN
INTERNATIONAL SEMINAR ON GRAVITATION AND COSMOLOGY, Joao
Pessoa, 2008, to be published in the special issue of
Int. J. Mod. Phys. A, Editors: V.B.Bezerra, V.M.Mostepanenko
and C.Romero}. \
Quantum gravity is an important and to great extent unsolved 
problem. There are many different approaches to the quantization 
of the metric field, both perturbative and non-perturbative. 
The current situation in the perturbative quantum gravity is
characterized by a number of different models, some of them 
well elaborated but no one perfect nor mathematically neither 
phenomenologically, mainly because there are no theoretically 
derived observables which can be experimentally measured. A 
very interesting one is an effective approach which separates 
the low-energy quantum effects from the UV sector. In this 
way one can calculate quantities which are potentially 
relevant for establishing certain universal features of 
quantum gravity. In this presentation we give a polemic 
consideration of the effective approach to the infrared 
quantum gravity. We question the validity of the recent 
results in this area and also discuss how one can check the 
alleged universality of the effective approach. 
\end{quotation}
\vskip 1mm

\vskip 1mm


\vskip 8mm

\section{Introduction}

General Relativity and Quantum Field Theory are very 
successful theories of the most fundamental phenomena
from the scale of elementary particles up to the scale 
of the whole Universe.
The problem of creating the quantum theory of gravity, that 
means a quantization of the metric field, is in the agenda of 
theoretical physicists for more than 70 years~\cite{Bronstein} 
and nowadays 
we may be proud of many important achievements and ideas. 
However, there is no real solution of the problem yet, in 
part due to the theoretical difficulties but also because 
there are no experimental data which could help to distinguish
more successful theories and models from the other ones. In 
this situation the advent of an idea of effective approach 
to the infrared quantum gravity,~\cite{don94} (we abbreviate 
it as IRQG in what follows) is very welcome, for it paves 
the way for the use of the conventional Feynman diagram 
technique to derive some concrete observables, such as 
quantum corrections to the Newton potential.  

Despite the idea of such calculations was not really 
new~\cite{Iwa}, the proposal of Ref.~\cite{don94} has a 
few very attractive features, namely: 

1) Universality of the IRQG. This means that one does not need 
to know what is the fundamental theory of quantum gravity at 
high energy scale in order to calculate at least some of 
the low-energy observables. The higher derivative terms 
and corrections which are inevitable in the framework 
of semiclassical approach or in string theory, do not 
concern the IR sector, which is completely related to 
the Einstein-Hilbert term in the gravitational action. 

2) The massive modes (degrees of freedom) of the gravitational 
field which may be present in an unknown fundamental theory, 
do decouple at low energies and we arrive at the quantum 
corrections which are due to the diagrams with only 
graviton and (massive) matter field internal lines. 
The same is true for the possible high derivative 
interactions of the gravitational degrees of freedom,
which are Planck suppressed and therefore irrelevant at low 
energy scale. 

After the papers~\cite{don94} were published, people readily 
noticed that, technically, the calculations presented there 
were not perfect. In particular, some diagrams were 
calculated incorrectly~\cite{IRQG-2set} and some relevant 
ones were missed. The last point was explained using the 
analog model based on scalar QED~\cite{SQED}. 
Starting from the set of diagrams presented in the 
original publication~\cite{don94}, one ends up with 
the quantum correction to the Newton potential 
which depends on an arbitrary choice of the gauge fixing 
condition, making all calculation senseless. The gauge 
fixing invariance of the amplitudes is restored when 
all relevant diagrams, in the given loop order, are 
taken into account. This property has been also verified 
for IRQG, but unfortunately the corresponding work was 
never published~\cite{AFirme}. 

Several calculations of the IRQG corrections to the 
Newton potential has been presented in the 
meanwhile~\cite{DaMazz,IRQG-2set,KKh,DHB}. It is 
stated that there is a correct result for 
the IRQG improved Newton potential~\cite{DHB} (there is, 
however, a divergence with the alternative calculation 
in Ref.~~\cite{KKh}. This result for the quantum correction 
includes $r^{-2}$ and  
$r^{-3}$--proportional terms which come from two different 
sets of diagrams: the ones which have {\it only} massless 
(graviton) internal lines give $L$-type contributions, which 
lead to the  $r^{-3}$ terms in the potential~\cite{don94}, 
while more complicated $Q$-type 
contributions come from the diagrams with 
two types of internal lines: both massless 
gravitons and massive (e.g. scalar) propagators are there. 
This kind of diagrams lead to both $r^{-2}$ and  $r^{-3}$ 
type corrections. Let us mention that in one of the cited
above calculations~\cite{DaMazz} the subject of quantization 
was only a metric field, but not the matter sources, which 
have the form of point particles. In this case the result 
for improved Newton potential is quite different, because 
only the $L$-type contributions
are present. It is important that such ``reduced'' quantum
corrections are gauge-fixing independent by themselves, 
without taking the $Q$-type contributions into account. 

The purpose of the present communication is to critically 
discuss the main points of the standard IRQG approach. In 
particular, we present some arguments questioning the 
universality of the IRQG. In our opinion this issue is not 
completely clear and requires an explicit verification. 
Furthermore, we argue that only the easy to calculate 
$L$-type contributions have reasonable physical 
interpretation while the complicated $Q$-type 
ones should be perhaps disregarded. 
We present only qualitative arguments and postpone 
the corresponding calculations for the next occasion. 

The paper is organized as follows. In the next section we
discuss the choice of the diagrams and in section 3 
consider the implications for the universality of the 
IRQG and how it can be, in principle, checked. In 
section 4 we draw our conclusions.

\section{Which Diagrams Must Be Taken And Which Not?}

In the first paper on IRQG~\cite{don94} there was a following 
important result: the sum of the corresponding terms from the 
$Q$-type diagrams reproduce the post-Newtonian limit of 
classical GR. It was correctly stressed that any other output 
would make the whole scheme of IRQG being rather suspicious, 
because the mentioned post-Newtonian correction is one of 
very successful tests of GR. However, the consequent analysis 
of quantum corrections has shown that the successful output of 
the original calculations was just a result of an unintentional 
wrong use of Feynman diagrams~\cite{IRQG-2set}. 
In particular, the most complicated diagrams were 
disregarded in~\cite{don94}. 
As we have already mentioned above, without the full 
set of diagrams the quantum corrections are gauge fixing 
dependent and the whole calculation has no sense~\cite{SQED}. 
Finally, in the mathematically correct result~\cite{DHB} 
there is no correspondence to the conventional post-Newtonian 
limit and also there is some intrinsic arbitrariness of the 
quantum corrections. One can think that something may be 
wrong either in the definition of the quantum theory or 
in the interpretation of the results. The immediate 
conclusion can be, for instance, that gravity should 
not be quantized at all. Fortunately, as we shall see 
in a moment, we do not need to go so far. 

Let us look at the situation from another viewpoint. 
We can remember, once again, that the $Q$-type diagrams 
consist of the loops with mixed internal lines content, 
that means there are, at the same time, massless graviton 
and massive matter lines. As a model for the matter it is
usually taken a massive scalar field. This model works 
well for the one-graviton exchange between the two 
masses $m_1$ and $m_2$, e.g., it is perfectly producing 
a Newton law in the nonrelativistic regime. So, it looks 
natural to go beyond the tree-level approximation and try 
to evaluate loop corrections~\cite{Iwa,don94}. 

In the path integral interpretation of Quantum Field Theory 
the presence of massive matter lines in the {\it internal} 
part of Feynman diagrams means that the matter field is a 
subject of functional integration, and it means that 
this field must be quantized. 
In order to see this one can compare the two possible 
forms of the generating functionals of the Green functions  
\beq 
Z_1 [J^{\mu\nu}, \,{\cal J}]
= \int {\cal D} g_{\mu\nu} \,  {\cal D} \Phi
\,\,e^{i S[g_{\mu\nu}, \,\Phi] + i g_{\mu\nu} J^{\mu\nu}
+ i \Phi {\cal J}}
\label{Z1}
\eeq
and 
\beq 
Z_2 [J^{\mu\nu}, \,{\cal J}] = \int {\cal D} g_{\mu\nu} 
\,\,e^{i S[g_{\mu\nu}, \,\Phi] + i g_{\mu\nu} J^{\mu\nu}
+ i \Phi {\cal J}}
\label{Z2}
\eeq
In both expressions we meet a functional integration over 
the metric, but in the first case there is an additional 
integral over the matter field $\Phi$. Only this 
integration enables one to have propagators of this field 
in the internal part of the loops. At the same time, the 
presence of the source ${\cal J}$ for the  field $\Phi$ 
lets us to have the diagrams with {\it external} lines of  
field $\Phi$ in both cases. From the other side, the 
functional integration over the field $\Phi$ is possible only 
if this field is an elementary quantum object~\cite{Popov} 
and not a classical source or a composite field, because 
the last should be treated in a different way. 

The first option (\ref{Z1}) looks more general, the 
calculations of Ref.~~\cite{don94} and most of the previous 
and consequent calculations (except Ref.~~\cite{DaMazz}) of 
the quantum corrections to the Newton potential were based 
on (\ref{Z1}), with  $\Phi$ being a scalar field. 
The bad news for this approach is that, after all, the 
macroscopic bodies (e.g. planets or stars, satellites etc) 
which take part in the phenomenologically relevant 
gravitational interactions are not made from a scalar 
field. Much on the contrary, they do consist from a 
baryonic matter, that means interacting protons, neutrons 
and electrons. These particles are not elementary  
(except electron) and none of them may be 
properly described by a scalar field. Of course, nucleons 
consist from quarks and gluons, so one may think to replace 
the scalar field by the spinor one and try to obtain the 
quantum gravity corrections taking, e.g., mixed graviton-quark 
diagrams. However, this would not be a right step, because 
quarks are not free particles. One of the manifestation of 
this fact is that, e.g., the total mass of the $u$, 
${\bar u}$ and $d$ quarks inside the proton is essentially 
smaller than the mass of the whole proton. Therefore, if 
we calculate such (even tree-level) diagrams with quarks 
we have no chance to arrive at the correct result. The 
same is true for the protons and neutrons, which are 
not free but instead interact within the nucleus. 
After all, the $Q$-type diagrams imply the quantization 
of macroscopic bodies, e.g. of the Earth, Moon or Mercurio. 
It seems to us that such quantization is something odd 
with respect to the principles of quantum theory and 
therefore it should be better avoided. Finally, the 
most correct approach is to treat massive sources 
correctly, that means to regard them as massive 
macroscopic bodies which should not be quantized. 

Finally, we arrive at the conclusion that the ``correct'' 
set of diagrams 
includes {\it only} the $L$-type ones. Indeed this point 
of view was already presented in Ref.~~\cite{DaMazz}, where 
the corresponding quantum corrections were calculated 
through the functional method for the massive source 
which consists of the point particles. The analysis of 
these calculations shows that one can chose any other 
form of external source (e.g., scalar or fermion field) 
without 
changing the result. Also, the same quantum corrections
can be probably obtained through the $L$-type Feynman 
diagrams. These diagrams, with only graviton internal 
lines are in fact easy to evaluate. In fact, the 
quantum contributions coming from the massless diagrams 
are always easy to evaluate, because the IR non-local terms
are dual to the UV divergences, which can be obtained, 
e.g., through the Schwinger-DeWitt technique. The 
resulting quantum corrections will be only of the $r^{-3}$ 
type and hence they are not very relevant from the 
phenomenological viewpoint~\cite{don94}. However, the 
very fact we can separate the IRQG effects from the 
UV sector remains very interesting by itself. 

\section{Is The IRQG Really Universal?}

The $r^{-3}$ type quantum corrections to the Newton 
potential are very small, but the possibility to 
evaluate them in a unique and consistent way is
exciting, especially taking into account the existing
variety of the Quantum Gravity models. However, let us
inspect whether the universality of the IRQG is a certain
thing or it is only a hypothesis which has to be verified. 

One can distinguish the possible corrections to the 
Einstein-Hilbert action in the following two kinds of 
theories: (super)string theory and the quantum theory of 
matter fields on curved gravitational background. In the 
first case the action of gravity is the low energy effective 
action which emerges after we quantize the fundamental 
object - a (super)string. Within a standard Polyakov 
approach~\cite{Pol81} this effective action has a form 
of an expansion in the parameter $\al^\prime$. At the 
first order one meets the Einstein-Hilbert action with 
additional dilaton field and at the next orders there 
are higher derivative corrections to this action. After 
one obtains the low energy string effective action, the 
two additional operations are executed. From one side, 
one needs to compactify extra dimensions. Furthermore,
it is customary to perform the Zweibach reparametrization 
of the metric~\cite{zwei} in order to make the higher 
orders in 
$\al^\prime$ terms free of the high derivative ghosts. 
Indeed, this operation is very ambiguous~\cite{torsi}
but one can hope this feature disappears when using an 
exact result which is nonperturbative in $\al^\prime$. 
From the IRQG viewpoint, the Zweibach 
transformation means the graviton propagator can be 
derived from the Einstein-Hilbert action alone, without 
taking the higher derivative (that means higher order in 
$\al^\prime$) corrections into account. These corrections
show up only in the vertices and are suppressed by the 
Planck scale. Therefore, in this case the scheme of IRQG
works perfectly well and we have the desired universality
of the quantum corrections. 

One has to keep in mind that the known low energy 
quantum effects are described not by the string theory 
but by the quantum field theories, such as the Standard 
Model of particle physics or its generalizations. 
Following this pattern, in 
the presence of a gravitational field one has to use the 
formalism of quantum field theory (QFT) in curved space
(see, e.g. Refs.~~\cite{birdav,book} for the introduction 
and Ref.~~\cite{PoImpo} for a recent review). The bad news 
for IRQG is that the standard formalism of QFT in curved 
space implies the formulation of a classical action of 
external metric and that this classical action includes
dynamical higher derivative terms, like the square of 
the Weyl tensor $C^2=C_{\mu\nu\al\be}C^{\mu\nu\al\be}$
and of the scalar curvature,
\beq
S_t = \int d^4x\sqrt{-g}\,\left\{
\,- \frac{1}{16\pi G}\,\big(R+2\La\big) 
- \frac{1}{2\la}\,C^2 + \frac{\om}{3\la}\,R^2\right\}\,.
\label{total ac}
\eeq
The fourth derivative terms are necessary for 
renormalizability of the theory and do not lead to the 
problem with unphysical massive spin-2 ghosts because 
in this approach we do not need to consider the $S$-matrix 
for the gravitational excitations. Therefore, despite 
the procedure of metric reparametrization similar to 
the one used in string theory~\cite{zwei} is possible,
there is no reason to apply it, especially in view
of serious ambiguities which follow from this procedure. 

In the model of IRQG, however, we have to quantize the 
metric and therefore meet the usual problem of unphysical
ghosts. In this situation one can apply the same logic as 
people use in the 
string theory: first calculate quantum corrections to 
any interesting physical observable starting from the 
higher derivative action (\ref{total ac}) or from its 
superrenormalizable generalizations~\cite{highderi} and 
then perform the transition to the ``observable'' metric 
via the Zweibach reparametrization. It is obvious that 
the higher derivative quantum gravity becomes a perfectly
consistent theory within this approach. However, by the 
end of the day we meet the mentioned ambiguity 
related to the transformation~\cite{zwei}. 

Now, let us come back to the IRQG and see what are the 
implications of the general quantum gravity situation in 
this case. The low-energy quantum corrections depend on 
the following three elements: gravitational propagator, 
vertices of gravitational self-interaction and 
vertices of gravitational interaction with matter. 
The last does not depend on the presence of the higher 
derivative terms in the gravitational action. Next, 
looking at the action (\ref{total ac}) it is clear that 
contributions of the higher derivative terms to the 
vertices of gravitational self-interaction are suppressed,
in the IR, by the ratio of typical energy of the process
to the Planck mass. Therefore the only potentially doubtful 
element of the IRQG technique is the propagator of the 
gravitational field. 

Here we meet the following dillema: since the 
graviton propagator has to be derived from the total 
action, how can we know that the low-energy sector of the 
propagator comes from the Einstein term and not from 
the fourth derivative terms? The standard argument in 
favor of universality of GR as IRQG is covariance which 
means $\sqrt{-g}R$-term is the unique second derivative 
covariant term. However, the whole action (\ref{total ac})
is covariant and this does not mean that the relevant 
at low energies part of the propagator has tensor 
structure of the Einstein term and does not depend 
on the fourth derivative ($\sqrt{-g}C^2$ and $\sqrt{-g}R^2$) 
terms. 

It is well known that the propagator of the gravitational 
field depends on the gauge fixing conditions and, in case 
of the higher derivative quantum gravity, there are more 
degrees of freedom in the propagator than in the theory 
based on General Relativity. The theory (\ref{total ac}) 
can be seen as describing the interaction of two different 
particles: massless graviton and massive (including spin-2) 
ghosts (we do not discuss the problem of unitarity here). 
Therefore, the relevant Feynman diagrams include the loops 
of massive components of the metric and also the mixed 
loops with both massless and massive internal lines. 
Therefore the problem of the quantum calculations in 
this theory is technically similar to the one addressed 
in Refs.~~\cite{don94,Iwa,SQED,IRQG-2set,DHB,ClassQuant,KKh}.
One can consider the propagator of higher derivative 
gravity as an algebraic sum of the graviton propagator 
and the propagator of a massive unphysical 
ghost~\cite{stelle,julton}. If we consider the diagrams 
which do contribute to, for example, quantum corrections 
to the Newton potential, these extra degrees of freedom 
will show up in the mixed loops, with both graviton and 
ghost internal lines. Using the method of~\cite{KKh,DHB},
we can reduce these diagrams to the ones without ghost 
internal lines, but they remain relevant and therefore 
can contribute to the corrections to the Newton potential.
It looks like the cancelation of the corresponding quantum 
contributions is required to provide the irrelevance of 
the higher derivative terms. In other words, we have to 
check that the contributions of the higher derivative 
ghosts really decouple at low energies~\cite{Gauss}. 

Finally, we can see that simple qualitative considerations 
do not ensure the universality of the IRQG and one needs an 
explicit verification. Such verification can not be performed 
in the framework of the quantum GR and requires the analysis 
of some more general theory, e.g., the one based on the action 
(\ref{total ac}) or some its alternatives~\cite{highderi}. 
This means one has to start from one of the more general 
actions and calculate its low-energy prediction. If the 
main hypothesis of Ref.~~\cite{don94} is right, the effect 
of all terms except the ones of $\sqrt{-g}R$-term will be 
negligible in the IR.   Practical realization of such 
calculations is possible, at least in the framework of 
the fourth derivative model (\ref{total ac}), where we 
have an extensive experience of loop 
calculations~\cite{julton,frts82,avrbar,Gauss}. However, 
what is actually requested for the IRQG purposes is 
much more complicated, because one has to go beyond 
the conventional Minimal Subtraction renormalization 
scheme, because one need to extract some relevant 
information about the finite part of the effective 
action.

\section{Conclusions}

We have considered some qualitative arguments 
concerning the model of IRQG. There are serious arguments 
in favor of restricting the set of relevant diagrams such 
that the massive matter source is not quantized. Also, if 
we quantize only metric and look for the corresponding 
logarithmic corrections to the amplitudes, it is not 
obvious that the low-energy limit of the theory 
possesses the alleged universality. In order to ensure 
that this nice property really takes place one has to 
start from the theory more general than the GR, derive 
some observable in this framework and compare it to the 
one obtained within the quantum GR. The most important 
example of a more general gravitational theory is 
the model based on the action (\ref{total ac}),
because this form of the action is dictated by the 
renormalizability of the semiclassical theory. 

On the top of that, there is a possibility to define both 
semiclassical theory and perturbative quantum gravity with 
an additional procedure of metric reparametrization, which 
must be performed {\it after} the quantum corrections are 
calculated, in the same way as it is done in the string 
theory~\cite{zwei}. In this way one can construct a theory 
of quantum metric which would be both renormalizable and 
free of high derivative ghosts. However, there is a serious 
price to pay. The disadvantages of this scheme are the 
noncanonical use of the quantum field theory procedure 
and, also, a vast ambiguity in the quantum corrections. 

\section*{Acknowledgments}
Author is grateful to Blazenka Melic and Guilherme de 
Berredo Peixoto for useful discussions.
His work was supported by CNPq, FAPEMIG, FAPES and ICTP.


\end{document}